\documentclass{aa}  

\usepackage{graphicx}
\usepackage{txfonts}
\usepackage{natbib}
\usepackage{float}
\usepackage{enumitem}
\usepackage{subcaption}
\usepackage{mwe}
\usepackage{bm}
\usepackage{multirow}
\usepackage[switch]{lineno}
\usepackage{hyperref}
\hypersetup{
  colorlinks=true,   
  allcolors=blue
}

\newcommand{\maspy}{$\mathrm{mas~yr^{-1}}$}
\newcommand{\uaspy}{$\mu\mathrm{as~yr^{-1}}$}
\newcommand{\mjypb}{$\mathrm{mJy~beam^{-1}}$}

\newcommand{\psr}{PSR~J2222$-$0137}
\newcommand{\ibca}{FIRST~J222112$-$012806}
\newcommand{\ibcb}{FIRST~J222201$-$013236}
\newcommand{\agna}{ICRF~J221852.0$-$033536}
\newcommand{\agnb}{ICRF~J221947.2$-$005132}
\newcommand{\agnc}{ICRF~J222646.5$+$005211}
\newcommand{\blz}{ICRF~J214805.4$+$065738}
\newcommand{\multilinecomment}[1]{}

\begin{document}

   \title{A millisecond pulsar position determined \\to 0.2\,mas precision with VLBI}

   \author{Hao Ding
          \inst{1}\fnmsep\thanks{EACOA Fellow}
          \and
          Adam T. Deller\inst{2}
          \and
          Paulo C. C. Freire\inst{3}
          \and
          Leonid Petrov\inst{4}
          }

   \institute{Mizusawa VLBI Observatory, National Astronomical Observatory of Japan, 2-12 Hoshigaoka-cho, Mizusawa, Oshu, Iwate 023-0861, Japan\\
              \email{hdingastro@hotmail.com}
              \and
    Centre for Astrophysics \& Supercomputing, Swinburne University of Technology, PO Box 218, Hawthorn, Victoria 3122, Australia
         \and
    Max-Planck-Institut f$\ddot u$r Radioastronomie, Auf dem H$\ddot u$gel 69, D-53121 Bonn, Germany
    \and
    NASA Goddard Space Flight Center Code 61A, 8800 Greenbelt Rd, Greenbelt, 20771 MD, USA
             }

   \date{}

 
  \abstract
   {Precise millisecond pulsar (MSP) positions determined with very long baseline interferometry (VLBI) hold the key to building the connection between the kinematic and dynamic reference frames respectively used by VLBI and pulsar timing. The frame connection would provide an important pathway to examining the planetary ephemerides used in pulsar timing, and potentially enhancing the sensitivities of pulsar timing arrays used to detect stochastic gravitational-wave background at nano-Hz regime.}
   {We aim at significantly improving the VLBI-based MSP position from its current $\gtrsim1$\,mas precision level by reducing the two dominant components in the positional uncertainty --- the propagation-related uncertainty and the uncertainty resulting from the frequency-dependent core shifts of the reference sources.}
   {We introduce a new differential astrometry strategy of using multiple calibrators observed at several widely separated frequencies, which we call PINPT (Phase-screen Interpolation plus frequeNcy-dePendent core shifT correction; read as "pinpoint") for brevity. The strategy allows determination of the core-shift and mitigates the impact of residual delay in the atmosphere. We implemented the strategy on \psr, an MSP well constrained astrometrically with VLBI and pulsar timing.}
   {Using the PINPT strategy, we determined core shifts for 4 AGNs around \psr, and derived a VLBI-based pulsar position with uncertainty of 0.17\,mas and 0.32\,mas in right ascension and declination, respectively, approaching the uncertainty level of the best-determined timing-based MSP positions. Additionally, incorporating the new observations into historical ones, we refined the pulsar proper motion and the parallax-based distance to the $\lesssim10$\,\uaspy\ level and the sub-pc level, respectively. 
   }
   {The realization of the PINPT strategy promises a factor-of-5 positional precision enhancement (over conventional VLBI astrometry) for all kinds of compact radio sources observed at $\lesssim2$\,GHz, including most fast radio bursts.}


   \maketitle
%

\section{Introduction}
\label{sec:intro}

Millisecond pulsars (MSPs) refer to recycled fast-spinning neutron stars, which exhibit unparalleled spin stability compared to other pulsars \citep{Hobbs10}. Using the pulsar timing technique \citep[e.g.][]{Detweiler79} that time and model the pulse arrival times, astronomers have delivered the most stringent tests of gravitational theories with MSPs \citep[e.g.][]{Freire24}. 
Collectively, an array of MSPs scattered across the sky, as known as a pulsar timing array (PTA), can be used to directly probe the stochastic gravitational-wave background (GWB) at the nHz regime \citep{Sazhin78}.

Recent years have seen the major PTA consortia closing in on achieving high-significance detections of a homogeneous GWB \citep{Agazie23a,Reardon23a,Antoniadis23a,Xu23}.
Despite the breakthrough, to deepen our understanding of the sources of the GWB still requires continuous improvement of the PTA sensitivities.
The optimal strategy to sustain PTA sensitivity enhancement is to regularly add new MSPs to the PTAs \citep{Siemens13}, as has been adopted by the MPTA \citep{Miles23}. However, it is generally difficult to quantify the red timing ``noises'' (in which the GWB signal resides) for a shortly timed ($\lesssim3000$ days) MSP; one way to overcome this difficulty is to incorporate independent astrometric measurements (i.e., sky position, proper motion and parallax) into the inference of timing parameters \citep{Madison13}.

The very long baseline interferometry (VLBI) technique can provide precise, robust and model-independent astrometric measurements for MSPs, and takes much shorter time to achieve a certain astrometric precision, as compared to the astrometric determination made with pulsar timing \citep[e.g.][]{Brisken02,Chatterjee09,Deller19}. 
Therefore, incorporating precise VLBI astrometric measurements into timing analysis of MSPs plays an essential role in testing gravitational theories \citep[e.g.][]{Deller08,Kramer21a}, and may substantially enhance the PTA sensitivities \citep{Madison13}.

However, the incorporation is technically challenging.
Firstly, incorporating precise VLBI proper motion and parallax into timing analysis can be limited by potential temporal structure evolution of the reference sources used in VLBI astrometry.
Secondly, incorporating a VLBI pulsar position into timing analysis hinges on a good understanding of the transformation between the two distinct kinds of reference systems used by VLBI astrometry and pulsar timing. Though both VLBI astrometry and pulsar timing are usually presented with respect to the barycenter of the Solar System, VLBI astrometry is conducted in the kinematic reference frame established with remote AGNs quasi-static on the sky (hence being robust to inaccurate planetary ephemerides), while pulsar timing studies are carried out in the dynamic reference frame that requires reliable planetary ephemerides to convert Earth-based pulse arrival times to the barycenter of the solar system. 
The dynamic coordinate is anchored to a kinematic coordinate system 
through observations of common objects, for instance, differential
observations of asteroids with respect to stars. VLBI observations
of MSPs with respect to AGNs will allow us to determine a rotation
of the dynamic coordinate system defined by planetary ephemerides
with respect to the inertial coordinate system (based on VLBI
observations of AGNs) \citep{Madison13,Wang17,Liu23a}.

The residuals in pulsar positions from VLBI and timing observations 
after a subtraction of the rotation will allow us to provide an independent assessment of pulsar timing errors and validate the PTA error model. We consider that question as a matter of great importance because a claim that a PTA has detected GWB is based upon a model of pulsar timing errors.

So far, planetary-ephemeris-dependent frame rotation remains poorly constrained, mainly limited by relatively large ($\gtrsim1$\,mas) VLBI position uncertainties of MSPs \citep[e.g.][]{Liu23a}, as compared to the $\lesssim0.2$\,mas timing position uncertainties for the best-timed MSPs \citep[e.g.][]{Perera19}.
According to \citet{Liu23}, 50 VLBI-based MSP positions with the current precision level would constrain the frame rotation to the 0.3\,mas level; this constraint would be improved to the 0.1\,mas level, if VLBI could determine MSP positions to 0.3\,mas precision.
Therefore, to reduce the VLBI position uncertainties of MSPs holds the key to building the planetary-ephemeris-dependent frame tie, examining the quality of planetary ephemerides, and hence facilitating the quantification of the timing noises resulting from inaccurate planetary ephemerides. 

In this paper, we introduced and tested a novel method to significantly improve the precision of pulsar VLBI positions.
Throughout the paper, uncertainties are provided at 68\% confidence, unless otherwise stated; mathematical expressions (including the subscripts and superscripts) defined anywhere are universally valid.

\section{A novel observing strategy \& test observations}
\label{sec:PINPT_and_obs}

\subsection{The PINPT strategy}
\label{subsec:PINPT}

As radio pulsars are generally faint ($\sim1$\,mJy at 1.4\,GHz, and with steep spectra), the standard approach for directly determining positions in the quasi-inertial VLBI frame involving the measurement of group delays is not feasible. Instead, pulsar absolute positions have been determined using differential astrometry with respect to relatively bright reference sources, which are normally AGNs that are not point-like. 
Since the position of a suitable nearby AGN can be determined using standard absolute astrometry techniques, such relative position measurements allow a connection of the pulsar position into the quasi-inertial frame. However, standard absolute astrometry techniques do not account for any structure in the AGN, and so the position extracted for these sources is effectively that of the peak brightness in the image.
The brightest spot in the 2D brightness distribution (or simply image) of an AGN also serves as the reference point of differential astrometry, which is usually the optically thick jet core (as long as the AGN is not flaring). As the AGN images normally vary with observing frequency $\nu$, the reference point (or the jet core) evolves with $\nu$ as well; additionally, frequency-dependent image models of the reference sources are required for pulsar astrometry.

Three main error sources contribute to the error budget of the absolute pulsar position (see Sect.~3.2 of \citealp{Ding20}): {\bf i)} the uncertainty in the absolute position of the primary phase calibrator (derived and registered in the Radio Fundamental Catalogue\footnote{\label{footnote:rfc}\url{http://astrogeo.org/}}, or the ICRF3 catalog, \citealp{Charlot20}), {\bf ii)} the unknown frequency-dependent core shifts (hereafter simply referred to as core shifts) \citep[e.g.][]{Bartel86,Lobanov98} of the reference sources (or more generally, the frequency evolution of reference source structures), and {\bf iii)} differences in the line of sight propagation delay between the direction of the calibrator source (where it has been solved) and the direction of the target.
At L band, core shifts amounting to $\sim1.2$\,mas \citep{Sokolovsky11} usually dominate the error budget. Second to that, the propagation-related systematic error of the absolute pulsar position is also prominent, given the relatively large ($\gtrsim1$\,deg) separation between the pulsar and its primary phase calibrator.

To suppress the aforementioned positional uncertainties, we designed a special observing strategy of differential astrometry, which extends the core-shift-determining method pioneered by \citet{Voitsik18}, and combines the method with the Multi-View (referring specifically to {\it 2D interpolation} throughout this paper) strategy \citep[e.g.][]{Rioja17,Hyland23}.
The proposed observing strategy, referred to as the PINPT (Phase-screen Interpolation plus frequeNcy-dePendent core shifT correction; read as "pinpoint") strategy, requires a group of $\lesssim6$ observations for absolute position determination: $\lesssim3$ L-band Multi-View observations of the pulsar (pulsar sessions), and $\lesssim3$ core-shift-determining observations on nearby AGNs (that include the 3 calibrators used in the Multi-View session), each at different observing frequency $\nu$. 
Where suitable L-band in-beam calibrators are identified, the pulsar sessions can also be used as the L-band core-shift-determining session, hence reducing the required observing time.

Our observing strategy of multi-frequency observations with multiple phase calibrators has three advantages.
{\bf Firstly}, with the multi-frequency observations of AGNs, the core shifts of the AGNs would be well determined, which would significantly reduce the core-shift-related errors of the absolute pulsar position. 
{\bf Secondly}, the use of Multi-View strategy would remove propagation-related systematic errors to at least the first order \citep[e.g.][]{Ding20c}. 
{\bf Finally}, with three phase calibrators of the Multi-View setup, the pulsar position uncertainties due to uncertain reference source positions would drop by a factor of $\leq\sqrt{3}$ (see Sect.~4.4 of \citealp{Ding20c}), compared to using only one phase calibrator.

\subsection{Observations}
\label{subsec:obs}

To test the PINPT strategy, we ran four observing sessions in December 2021 and January 2022 using the Very Long Baseline Array (VLBA) on \psr, a millisecond pulsar well determined astrometrically by previous campaigns (\citealp{Deller13,Guo21}, hereafter referred to as D13 and G21). 
The observations, carried out under the project code BD244, include two 2-hr L-band Multi-View observations and two other 2-hr core-shift-determining observations at S/X band and Ku ($\sim15$\,GHz) band, respectively.
The first L-band observation was scheduled within 2 days of the S/X- and Ku-band observations (in order to minimize structure evolution of the phase calibrators), and served as both pulsar session and core-shift-determining session, whereas the second L-band observation was solely a pulsar session.

More specifically, in the L-band observations, 
a ``target pointing'' covers \psr\ and two in-beam (at L band) calibrators identified by D13, i.e., \ibca\ (J2221) and \ibcb\ (J2222); 
scans on this field were interleaved with scans on three brighter but off-beam AGNs (hereafter referred to as off-beam calibrators), i.e., \agna\ (J2218), \agnb\ (J2219), and \agnc\ (J2226), with a cycle time of 5 minutes.
In the S/X- and Ku-band observations, J2221 and the three off-beam  calibrators (hereafter referred to as the core-shift-probing AGNs) were observed alternately (by 5-min and 2-min cycles, respectively). 
All of the 4 core-shift-probing AGNs are selected to have displayed resolved jet-core radio features in VLBI data, which eases the core shift determination (see Sect.~\ref{subsubsec:CS_direction}).
Being fainter at higher observing frequencies, \psr\ was skipped in the S/X- and Ku-band observations.
The calibrator plans of the observations are displayed in Fig.~\ref{fig:calibrator_plan}. In addition to the phase calibrators relatively close to \psr\ on the sky, the bright blazar \blz, further away from the sky region of interest, was used as the fringe finder to correct instrumental delays and filter bandpass. The unresolved flux densities and the angular separations of the aforementioned sources can be found in Table~\ref{tab:sources}.

To precisely determine the core shifts, a wide variety of observing frequencies are required.
The S/X-band observation simultaneously covers two frequency bands, i.e., at around $\sim2.3$\,GHz and $\sim8.4$\,GHz.
Additionally, the L-band observations were designed to cover two separate frequency ranges centered around 1.44\,GHz and 1.76\,GHz, respectively.
Altogether, we sampled 5 distinct frequency ranges, for the purpose of core shift determination.
The observing configurations, the phase calibration strategies (see Sect.~\ref{sec:data_reduction}) and the purposes of the four BD244 sessions are summarized in Table~\ref{tab:obs_summary}.
The VLBA data prior to data reduction can be accessed with the project code BD244 at \url{https://data.nrao.edu/portal}.

\begin{table}[h]
     \centering
     \caption[]{\label{tab:obs_summary}
     Observing setups, phase calibration strategies, and the purposes of the four BD244 sessions.}
     {\renewcommand{\arraystretch}{1.4}
     \resizebox{\columnwidth}{!}{
    \begin{tabular}{c|c|c|c|c|c}
        \hline 
        \hline
        & \multirow{2}{*}{\shortstack{Project\\code}} & \multirow{2}{*}{BD244A} & \multirow{2}{*}{BD244B} & \multirow{2}{*}{BD244C} & \multirow{2}{*}{BD244D} \\
        & & & & & \\
        \hline

        \hline
      \multirow{9}{*}{\shortstack{Obs.\\setup}} & Epoch (MJD) & 59554 & 59555 & 59556 & 59595 \\
        \cline{2-6}
        
       & $\lambda$ (cm) & 21, 17 & 13, 4 & 2 & 21, 17 \\
       
       \cline{2-6}
       & \multirow{3}{*}{\shortstack{Target/\\in-beam\\calibrators}} & pulsar, & --- & --- & pulsar, \\
       &  & J2221, & J2221 & J2221 & J2221, \\
       & & J2222 & --- & --- & J2222 \\
       \cline{2-6}
       & \multirow{3}{*}{\shortstack{Out-of-beam\\ phase\\calibrators}} & J2219, & J2219, & J2219, & J2219, \\
       & & J2226, & J2226, & J2226, & J2226, \\
       & & J2218 & J2218 & J2218 & J2218 \\
       \cline{2-6}
      & $T_\mathrm{cycle}$ (min) & 5 & 5 & 2 & 5 \\
       \hline
       
     \multirow{6}{*}{\shortstack{Phase\\cal.\\strategy}} & \multirow{2}{*}{\shortstack{In-beam phase\\referencing?}} & \multirow{2}{*}{yes} & \multirow{2}{*}{---} & \multirow{2}{*}{---} & \multirow{2}{*}{yes} \\
       & & & & & \\
        \cline{2-6}
       & \multirow{2}{*}{\shortstack{Out-of-beam\\phase ref.?}} & \multirow{2}{*}{yes} & \multirow{2}{*}{yes} & \multirow{2}{*}{yes} & \multirow{2}{*}{---} \\
       & & & & & \\
        \cline{2-6}
       & \multirow{2}{*}{\shortstack{Multi-View\\calibration?}} & \multirow{2}{*}{yes} & \multirow{2}{*}{yes} & \multirow{2}{*}{---} & \multirow{2}{*}{---} \\
       & & & & & \\
       \hline

       \multirow{5}{*}{\shortstack{Purpose}} & \multirow{2}{*}{\shortstack{Target\\session?}} & \multirow{2}{*}{yes} & \multirow{2}{*}{---} & \multirow{2}{*}{---} & \multirow{2}{*}{yes} \\
       & & & & & \\
       \cline{2-6}
       & \multirow{3}{*}{\shortstack{Core-shift-\\determining\\sessions?}} & \multirow{3}{*}{yes} & \multirow{3}{*}{yes} & \multirow{3}{*}{yes} & \multirow{3}{*}{---} \\
       & & & & & \\
       & & & & & \\

        \hline
        
        \hline
    \end{tabular}
    }}
    \tablefoot{$\lambda$ and $T_\mathrm{cycle}$ refer to observing wavelength and the target-calibrator cycle time (see Sect.~\ref{subsec:obs}), respectively.
    The calibrator plan is plotted in Fig.~\ref{fig:calibrator_plan}, with the source information provided in Table~\ref{tab:sources} and Sect.~\ref{subsec:obs}.
    }
    \end{table}

\section{Data reduction \& Direct results}
\label{sec:data_reduction}

All data reduction was performed with the {\tt psrvlbireduce}\footnote{\label{footnote:psrvlbireduce}\url{https://github.com/dingswin/psrvlbireduce}} pipeline, which runs functions of {\tt AIPS} \citep{Greisen03} through {\tt ParselTongue} \citep{Kettenis06}, and images sources with {\tt DIFMAP} \citep{Shepherd94}. 
The data reduction follows a standard workflow described in \citet{Ding20}, except for the phase calibration.
Despite the total observing time of only 8 hours, the phase calibration of this work is sophisticated. There are two different procedures of phase calibration, which depend on the purpose of the session, i.e., whether it is target session or core-shift-determining session. 
As noted in Sect.~\ref{subsec:obs}, the first L-band observation serves as both target session and core-shift-determining session, therefore being reduced twice in two distinct procedures.
The phase calibration strategies applied to the four observing sessions are summarized in Table~\ref{tab:obs_summary} and described as follows.

\subsection{Target sessions}
\label{subsec:target_sessions}

For the data reduction of target sessions, we did not split the data by observing frequency into two (one at around 1.44\,GHz and the other at around 1.76\,GHz). By doing so, the average central frequency 1.6\,GHz agrees with previous astrometric campaigns of \psr\ (D13, G21), and the acquired image S/N (and hence the positional precision) is not lowered.
As mentioned in Sect.~\ref{subsec:obs}, the two L-band observations involve 5 phase calibrators, including off-beam  calibrators and two in-beam calibrators (also see Fig.~\ref{fig:calibrator_plan} and Table~\ref{tab:sources}). We implemented the phase calibration of \psr\ in two different ways, depending on the astrometric goal.

\subsubsection{In-beam astrometry}
\label{subsubsec:in_beam_astrometry}

Previous astrometric campaign of \psr\ was carried out between October 2010 and June 2012, spanning 1.7\,yr (D13, G21). The two new target sessions extend the astrometric time baseline by a factor of 6.6 to 11.3\,yr, 
promising higher astrometric precision (especially for proper motion).
To capitalize on the long time baseline, we phase-referenced \psr\ to the same reference source (i.e., J2222)
used in the previous campaign, following the data reduction procedure of D13 (including using the same image model of J2222). The updated results of in-beam astrometry is reported in Sect.~\ref{subsec:astrometric_inference}.

\subsubsection{Multi-View (2D interpolation)}
\label{subsubsec:2D_interpolation}

As one form of the Multi-View strategy, 2D interpolation uses $\geq3$ reference sources to derive the phase solution at the sky position of the target \citep[e.g.][]{Rioja17}, which can at least remove propagation-related systematic errors to the first order.
In order to determine precise absolute position of \psr, we applied Multi-View (as part of the PINPT strategy) with the off-beam  calibrators, all of which have well determined absolute positions\textsuperscript{\ref{footnote:rfc}}. 
We reiterate that realizing the PINPT strategy requires a combination of Multi-View session(s) and core-shift-determining session(s); the two kinds of sessions need to be arranged close to each other to minimize the effects of structure evolution.
The second L-band observation is $\approx40$ days apart from the core-shift-determining sessions. Therefore, Multi-View was applied to only the first L-band observation, but not the second one.

In practice, we realized the Multi-View in two different approaches described in Appendix~\ref{appendix:2D_interpolation}.
Both approaches consistently render the position $22^\mathrm{h}22^\mathrm{m}05\fs99997\!\pm\!0.1\,\mathrm{mas}$, $-01\degr37'15\farcs7825\!\pm\!0.2\,\mathrm{mas}$ for \psr\ at MJD~59554.0, and $22^\mathrm{h}22^\mathrm{m}01\fs373131\!\pm\!0.03\,\mathrm{mas}$, $-01\degr32'36\farcs97654\!\pm\!0.06\,\mathrm{mas}$ for J2222. 
For further examination, we also applied the two approaches of Multi-View to the X-band data (where the chance of phase wrap errors is much smaller than at L band), and confirmed with J2221 that the two approaches give almost identical positions. The availability of the J2222 position, as well as the J2221 position $22^\mathrm{h}21^\mathrm{m}12\fs680887\!\pm\!0.01\,\mathrm{mas}$, $-01\degr28'06\farcs30985\!\pm\!0.02\,\mathrm{mas}$, offers a distinct pathway to the absolute position of \psr\ (see Sect.~\ref{sec:results_and_implications}).
We note that the three positions presented above can only be considered relative positions with respect to the off-beam calibrators; and the positional uncertainties only include the statistical component related to random (thermal) noise in the VLBI image of \psr.

\subsection{Core-shift-determining sessions}
\label{subsec:CS_sessions}

The determination of AGN core shifts requires a wide coverage of observing frequencies. As mentioned in Sect.~\ref{subsec:obs}, the core-shift-determining sessions cover 5 frequency ranges. We first split the S/X-band data into S-band data and X-band data, and likewise split the wide-L-band data into two datasets --- one at $\sim1.44$\,GHz and the other at $\sim1.76$\,GHz. Thereby, we acquired altogether 5 datasets taken at 5 different observing wavelengths $\lambda$ --- 2\,cm, 4\,cm, 13\,cm, 17\,cm and 21\,cm.

From each of the 5 datasets, we measured the J2221 position with respect to each off-beam  calibrator by {\bf 1)} phase-referencing J2221 to each off-beam  calibrator, and {\bf 2)} dividing the phase-referenced J2221 data by its final image model obtained at $\lambda$, and {\bf 3)} fitting the J2221 position. 
In phase referencing, the final image model of an off-beam  calibrator at $\lambda$ was applied to the phase calibration of the off-beam  calibrator; subsequently, the acquired phase solution was applied to both the off-beam  calibrator and J2221.
To make the final image models of the off-beam  calibrators and J2221, we first applied 13\,cm models to the datasets of respective sources at all $\lambda$ during phase calibration (or self-calibration for J2221). In this way, we aligned the reference points of each core-shift-probing AGN across $\lambda$. Thereafter, we split out the aligned datasets, and remade the models of J2221 and off-beam  calibrators at each $\lambda$. 
The final image models of in-beam and off-beam  calibrators at all $\lambda$ were made publicly available\footnote{\label{footnote:image_models}\url{https://github.com/dingswin/calibrator_models_for_astrometry}}, to convenience the reproduction of our results.

\section{Astrometric parameters \& Core shifts}
\label{sec:inferences}

The data reduction described in Sect.~\ref{sec:data_reduction} produced {\bf a)} two pulsar positions measured with respect to J2222, {\bf b)} Multi-View positions of the pulsar, J2221, and J2222, and {\bf c)} $5\times3=15$ J2221 positions from the core-shift-determining sessions. The products {\bf a)} and {\bf c)} can provide stringent constraints on astrometric parameters and core shifts, respectively.

\subsection{Astrometric inference}
\label{subsec:astrometric_inference}

We added the two new pulsar positions measured with respect to J2222 to previous pulsar positions (measured with respect to J2222), and re-made the Bayesian inference as described in G21. The resultant astrometric parameters are provided in Table~\ref{tab:astrometric_results}. For comparison, the previous VLBI results reported by G21 are reproduced to Table~\ref{tab:astrometric_results}.
Unsurprisingly, the proper motion improves substantially by a factor of $\sim7$, while the parallax also being enhanced by $\approx14$\%, corresponding to a refined trigonometric distance of $268.6^{+1.0}_{-0.9}$\,pc for \psr. 
On the other hand, the $3\,\sigma$ discrepancy in $\mu_\delta$ reveals under-estimated systematic uncertainty in declination in previous works, likely due to vertical structure evolution of J2222, which has much larger impact on short ($\lesssim2$\,yr) timespan.

\begin{table}[htbp!]
     \centering
     \caption[]{\label{tab:astrometric_results}
     Enhanced astrometric results of \psr.
     
     }
     {\renewcommand{\arraystretch}{1.4}
     \resizebox{\columnwidth}{!}{
    \begin{tabular}{cccccc}
        \hline 
        \hline
        $\mu_\alpha$  & $\mu_\delta$ & $\varpi$ & $\Omega_\mathrm{asc}$ & $i$ & Reference \\
       (\maspy)   & (\maspy) & (mas) & (deg) & (deg) & --- \\ 
        \hline
        44.707(5) & $-5.403(12)$ & $3.723^{+0.013}_{-0.014}$ & 190(7)\tablefootmark{\,a} & 85.28(9)\tablefootmark{\,a} & this work \\
        44.70(4) & $-5.69(8)$ & $3.730^{+0.015}_{-0.016}$ & $189^{+19}_{-18}$ & 85.25(25) & G21 \\
        \hline
        
        \hline
    \end{tabular}
    }}
    \tablefoot{From left to right, the five parameters refer to the two components of proper motion, parallax, ascending node longitude, and orbital inclination. The reference epoch $t_\mathrm{ref}$ is MJD~55899. \\
    \tablefoottext{a}{The tighter constraints on the two orbital parameters almost entirely result from using more precise prior values obtained with pulsar timing (G21). Nonetheless, including the inference of $\Omega_\mathrm{asc}$ and $i$ is expected to improve the accuracy of the astrometric model.}
    }
    \end{table}

\subsection{Core shift determination}
\label{subsec:CS_determination}

Following the pioneering work by \citet{Voitsik18}, we developed new packages to infer the core shifts of the 4 core-shift-probing AGNs, from the 15 J2221 positions (with respect to three off-beam  calibrators at five $\lambda$) in a Bayesian style. 
The core shift inference requires three ingredients --- {\bf (1)} a prescription of systematic errors of the 15 J2221 positions, {\bf (2)} mathematical description of the core shifts, and {\bf (3)} an underlying mathematical relation between core shift and $\lambda$ (or observing frequency $\nu$). The three ingredients are addressed as follows.

\subsubsection{Systematic errors on measured J2221 positions}
\label{subsubsec:J2221_position_sysError}

Along with the 15 J2221 positions, their random errors due to noise in the J2221 images were also obtained with data reduction (described in Sect.~\ref{subsec:CS_sessions}). Additionally, atmospheric propagation effects would introduce systematic errors, which change with $\nu$ and the angular separation between J2221 and the reference source. We approached the systematic errors by 
\begin{equation}
\label{eq:J2221_sysError}
\frac{\sigma^\mathcal{S}_{ijk}}{1\,\mathrm{mas}}=\eta_\mathrm{EFAC}\cdot \eta_0 \cdot \frac{s_i}{1\,\mathrm{deg}} \cdot \hat{l}(\epsilon) \cdot \frac{\sigma_0(\nu_j)}{1\,\mathrm{mas}} \cdot \hat{\Theta}_{k} \,\,,
\end{equation}
where $\sigma^\mathcal{S}_{ijk}$ denote systematic errors; the subscript $i=\mathrm{J2218, J2219, J2226}$ specifies the reference source; the second subscript $j=1,2,3,4,5$ corresponds to one of the five observing frequencies (see Sect.~\ref{subsec:CS_sessions});
the last subscript $k=\alpha, \delta$ refers to right ascension (RA) or declination; $\eta_\mathrm{EFAC}$ is a scaling factor (for the estimated systematic uncertainty) to be determined with Bayesian analysis; $\eta_0$ is the initial scaling factor that brings $\sigma^\mathcal{S}_{ijk}$ to a reasonable value ($\sim1$\,mas) before inferring $\eta_\mathrm{EFAC}$, which eases the inference of $\eta_\mathrm{EFAC}$ at a later stage; $s_i$ represents the angular separation between J2221 and the reference source;  $\hat{\Theta}_k$ denotes the fractional synthesized beam size (projected to RA or declination); $\hat{l}$ and $\sigma_0$ stand for the two terms changing with antenna elevations $\epsilon$ and $\nu$, respectively. 
In this work, we set $\eta_0\equiv1$; the adopted $\hat{l}(\epsilon)$ and $\sigma_0(\nu)$ are described in Appendix~\ref{appendix:l0_and_sigma0}.

\subsubsection{The directions of core shifts}
\label{subsubsec:CS_direction}

Regarding the ingredient {\bf (2)}, we describe the core shift of the $i'$-th ($i'=\mathrm{J2218},\mathrm{J2219},\mathrm{J2221},\mathrm{J2226}$) AGN with two parameters --- the core shift magnitude $r_{i'}$, and the direction of core shift $\theta_{i'}$. 
As noted in \citet{Voitsik18}, it is difficult to infer both $r_{i'}$ and $\theta_{i'}$ for each of the 4 core-shift-probing AGNs from the 15 J2221 positions without prior constraints, which has been taken into account in design of the observations. As mentioned in Sect.~\ref{subsec:obs}, all of the core-shift-probing AGNs are selected to have displayed clear jet-core radio feature on VLBI scales. 
Following \citet{Voitsik18}, we assumed core shift directions are aligned with respective AGN jet directions, thereby gaining prior knowledge of the core shift directions with analysis of VLBI images of the AGNs. 
The determination of AGN jet directions is detailed in Appendix~\ref{appendixB:jet_direction_determination}.

\subsubsection{The relation between core shift and observing frequency}
\label{subsubsec:CS_freq_relation}

The relation between core shift and observing frequency is given by \citet{Konigl81} as $r \propto \nu^{-1/k_r}$, where the index $k_r$ equals to 1 when synchrotron self-absorption (that leads to the jet core radio emissions) is in equipartition \citep{Blandford79}. For the $i'$-th AGN, we {\it initially} adopted an equivalent formalism
\begin{equation}
\label{eq:r_freq_relation}
r_{i'} = r_{0i'}\left(\frac{\nu}{\nu_0}\right)^{-\beta_{i'}}
\end{equation}
as the relation between $r_{i'}$ and $\nu$, where $\beta_{i'}$ is a power index to be determined in Bayesian analysis; $v_0$ and $r_{0i'}$ refer to the reference frequency and the core shift magnitude at the reference frequency, respectively.

\begin{figure*}[h]
    \centering    \includegraphics[width=0.7\textwidth]{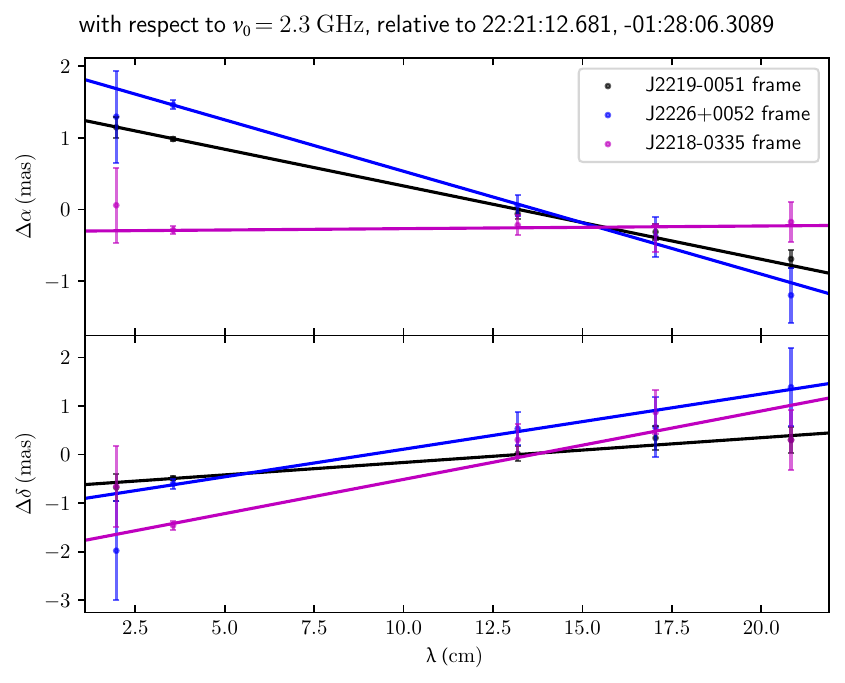}
    \caption{The jet core relative positions of \ibca\ measured with respect to the jet cores of three reference sources (i.e., \agna, \agnb, and \agnc) at five observing wavelengths. Systematic uncertainties calculated with Eq.~\ref{eq:J2221_sysError} have been included in the positional uncertainties.
    Models of the combined core shifts are derived assuming $\beta_{i'}\equiv1$ with Bayesian inference (see Appendix~\ref{appendix:Bayesian_inference_of_CSs}), and displayed with straight lines.}
    \label{fig:CS_results}
\end{figure*}

Using the Bayesian inference described in Appendix~\ref{appendix:Bayesian_inference_of_CSs}, we derive $\beta_{i'}$ along with other model parameters, which are provided in Table~\ref{tab:CS_results}. 
For the derivation, we choose $v_0=2.27$\,GHz, the median of the five central frequencies (see Sect.~\ref{subsec:CS_sessions} and Fig.~\ref{fig:CS_results}). The $r_{0i'}$ at a different $v_0$ can be calculated using Eq.~\ref{eq:r_freq_relation}.
When $\beta_{i'}$ are included in the inference, the reduced chi-square $\chi^2_\nu$ of inference is 1.9. Although $\beta_{i'}$ is significantly ($\gtrsim3\,\sigma$) determined for 2 of the 4 AGNs, it is consistent with 1 in all cases. In comparison, when performing the inference with all $\beta_{i'}$ fixed to 1, we acquired consistent results with generally higher precision. Moreover, the $\chi^2_\nu$ decreases to 1.2, suggesting $\beta_{i'}\equiv1$ as an appropriate assumption for this work. We adopt the results derived assuming $\beta_{i'}\equiv1$ for further analysis. 
The adopted core shift model is illustrated in Fig.~\ref{fig:CS_results}.

\section{An ultra-precise MSP position obtained with VLBI}
\label{sec:results_and_implications}

With $r_{0i'}$ and $\theta_{i'}$ determined for the three off-beam  calibrators and J2221, we calculated the core shifts of the 4 AGNs at 1.6\,GHz, and proceeded to derive the absolute position of \psr. The derivation was realized in three approaches. 
In the first approach, the pulsar position obtained with Multi-View (presented in Sect.~\ref{subsubsec:2D_interpolation}) was corrected for the core-shift refinements in the assumed positions of the off-beam  calibrators (as described by Eq.~\ref{eq:2D_interpolation_Delta_position_relation} in Appendix~\ref{appendix:deriving_abspos}).
We obtained the pulsar position $\alpha^\mathcal{G}_1=22^\mathrm{h}22^\mathrm{m}05\fs99993\!\pm\!0.17\,\mathrm{mas}$, $\delta^\mathcal{G}_1=-01\degr37'15\farcs7841\!\pm\!0.32\,\mathrm{mas}$ in the geocentric reference frame at the reference epoch $t_\mathrm{ref}$ of MJD~59554.0, which corresponds to $\alpha^\mathcal{B}_1 = 22^\mathrm{h}22^\mathrm{m}06\fs00015\!\pm\!0.17\,\mathrm{mas}$, $\delta^\mathcal{B}_1 = -01\degr37'15\farcs7827\!\pm\!0.32\,\mathrm{mas}$
with respect to the barycenter of the Solar System after removing the parallax effects.
Here, the positional uncertainty is the addition-in-quadrature of the statistical uncertainty (provided in Sect.~\ref{subsubsec:2D_interpolation}) and the $\Delta \vec{x}_\mathrm{target}$ uncertainty described in Appendix~\ref{appendix:deriving_abspos}.

In the second approach, we first acquired the J2221 position $\Vec{x}_\mathrm{J2221}$ at 1.6\,GHz by correcting the X-band J2221 position (obtained with Multi-View, see Sect.~\ref{subsubsec:2D_interpolation}) using the same process employed in the previous approach for the target source \psr.
Subsequently, we reprocessed all VLBA data (including the two new pulsar sessions and the historical ones) of \psr, phase-referencing \psr\ to J2221.
With the position series obtained from the data reduction, we inferred the reference pulsar position $\Vec{x}_\mathrm{psr}^\mathrm{\,J2221}$ (measured with respect to J2221) and its uncertainty at $t_\mathrm{ref}=\mathrm{MJD}~59554.0$ (along with other astrometric parameters) using the Bayesian analysis described in \citet{Ding23}.
Combining $\Vec{x}_\mathrm{J2221}$, the image model position $\Vec{x}_\mathrm{J2221}^\mathrm{\,model}$ of J2221, and $\Vec{x}_\mathrm{psr}^\mathrm{\,J2221}$, the absolute position of \psr\ was determined with Eq.~\ref{eq:abspos_via_IBC} to be $\alpha^\mathcal{B}_2 = 22^\mathrm{h}22^\mathrm{m}06\fs00015\!\pm\!0.19\,\mathrm{mas}$, $\delta^\mathcal{B}_2 = -01\degr37'15\farcs7825\!\pm\!0.42\,\mathrm{mas}$, where the error budget of the position is detailed in Appendix~\ref{subsec:pathway_thru_J2221} and Table~\ref{tab:pos_err_budget}.

The last approach is essentially the same as the second one, except that J2222 is used instead of J2221. Accordingly, the absolute pulsar position was calculated with the J2222 position $\Vec{x}_\mathrm{J2222}$ at 1.6\,GHz, the image model position $\Vec{x}_\mathrm{J2222}^\mathrm{\,model}$ of J2222, and the reference pulsar position $\Vec{x}_\mathrm{psr}^\mathrm{\,J2222}$ measured with respect to J2222. We obtained $\alpha^\mathcal{B}_3 = 22^\mathrm{h}22^\mathrm{m}06\fs00015\!\pm\!0.14\,\mathrm{mas}$, $\delta^\mathcal{B}_3 = -01\degr37'15\farcs7828\!\pm\!0.27\,\mathrm{mas}$.

The pulsar positions derived with the three approaches are summarized in the upper part of Table~\ref{tab:abs_pos}. The intermediate results leading to the three pulsar positions are provided in Table~\ref{tab:pos_err_budget}.
Though all the three approaches render highly consistent absolute pulsar positions, the first approach offers a snapshot (on the timescale of AGN structure evolution) localization (of \psr) independent from the previous astrometric observations of \psr, as opposed to the two other approaches (which essentially use the multi-frequency multi-source observations to perfect the position and structure of one of the in-beam calibrators, rather than the pulsar, at that snapshot in time, and then proceeds to perform standard differential astrometry using that frozen model of the astrometric reference source). 
Therefore, we report $\left(\alpha^\mathcal{B}_1,\delta^\mathcal{B}_1\right)$ at $t_\mathrm{ref}=\mathrm{MJD}~59554.0$ as the primary absolute position of \psr. 

\begin{table}[h]
     \centering
     \caption[]{\label{tab:abs_pos}
     Absolute pulsar positions with respect to the barycenter of the Solar System, derived with the PINPT strategy}
     {\renewcommand{\arraystretch}{1.5}
     \resizebox{\columnwidth}{!}{
    \begin{tabular}{c|c|c}
        \hline 
        \hline
          & $\alpha^\mathcal{B}$ & $\delta^\mathcal{B}$ \\
\cline{2-3}
            & \multicolumn{2}{c}{$t_\mathrm{ref}=\mathrm{MJD}~59554.0$ --- VLBI-only} \\
         \hline
        psr. appr. \tablefootmark{a} & $\mathbf{22^\mathrm{h}22^\mathrm{m}06\fs00015\!\pm\!0.17\,\mathrm{mas}}$ 
        & $\mathbf{-01\degr37'15\farcs7827\!\pm\!0.32\,\mathrm{mas}}$ \\
        J2221 appr. \tablefootmark{a} & $22^\mathrm{h}22^\mathrm{m}06\fs00015\!\pm\!0.19\,\mathrm{mas}$ 
        & $-01\degr37'15\farcs7825\!\pm\!0.42\,\mathrm{mas}$ \\
        J2222 appr. \tablefootmark{a} & $22^\mathrm{h}22^\mathrm{m}06\fs00015\!\pm\!0.14\,\mathrm{mas}$ 
        & $-01\degr37'15\farcs7828\!\pm\!0.27\,\mathrm{mas}$ \\

          \hline
        \hline
        
         & \multicolumn{2}{c}{$t_\mathrm{ref}=\mathrm{MJD}~55743$ --- VLBI vs timing} \\
        \hline
        J2221 appr. \tablefootmark{b} & $22^\mathrm{h}22^\mathrm{m}05\fs969005\!\pm\!0.13\,\mathrm{mas}$ 
        &  $\mathbf{-01\degr37'15\farcs72631\!\pm\!0.24\,\mathrm{mas}}$ \\
        J2222 appr. \tablefootmark{b} &  $\mathbf{22^\mathrm{h}22^\mathrm{m}05\fs969042\!\pm\!0.13\,\mathrm{mas}}$
        & $-01\degr37'15\farcs72646\!\pm\!0.21\,\mathrm{mas}$ \\
        \cline{2-3}
        DMX \tablefootmark{c} & $22^\mathrm{h}22^\mathrm{m}05\fs969046\!\pm\!0.18\,\mathrm{mas}$ 
        & $-01\degr37'15\farcs7257\!\pm\!0.5\,\mathrm{mas}$ \\
        non-DMX \tablefootmark{c} & $22^\mathrm{h}22^\mathrm{m}05\fs969071\!\pm\!0.06\,\mathrm{mas}$ 
        & $-01\degr37'15\farcs7267\!\pm\!0.1\,\mathrm{mas}$ \\
    
        \hline
        
        \hline
    \end{tabular}
    }}
    \tablefoot{The ``psr. appr.'', ``J2221 appr.'' and ``J2222 appr.'' refer to the three approaches (to the absolute pulsar position) described in Sect.~\ref{sec:results_and_implications}. 
  \\
    \tablefoottext{a}{Only the pulsar position at MJD~59554.0 obtained with the pulsar approach (shown in bold in the upper block) is free of additional errors due to structure evolution (of the reference sources), and is reported as the primary absolute position of \psr.}\\
    \tablefoottext{b}{For the sole purpose of comparing with the timing positions reported in G21, PINPT-based positions are derived for MJD~55743. To minimize the impact of structure evolution (of J2221 and J2222), the J2222-approach-based RA and the J2221-approach-based declination (shown in bold in the lower block) are adopted as the absolute pulsar position at MJD~55743 (see Sect.~\ref{subsec:comparison_VLBI_timing}).}\\ 
    \tablefoottext{c}{For readers' convenience, the latest published timing-based positions inferred without applying any VLBI prior (see Table~3 of G21) are listed here as ``DMX'' and ``non-DMX'' (see Sect.~\ref{subsec:comparison_VLBI_timing} for more explanations).}
    }
    \end{table}

\subsection{Comparison to timing positions}
\label{subsec:comparison_VLBI_timing}

A priori, we expect the position difference between the VLBI PINPT measurement, and that made by pulsar timing, to be consistent. With a sample size of 1, it is difficult to ascribe any discrepancy to one or the other of the measurement techniques, or to a systematic difference between dynamic and kinematic frames - but we can use the level of agreement to set a probabilistic upper limit on the error contribution from any of these three potential sources.
The most precise published timing-based position of \psr\ is reported in G21. In Table~3 of G21, three positions are provided for $t_\mathrm{ref}=\mathrm{MJD}~55743$, which include one position derived assuming a proper motion determined with VLBI, and two positions derived without using any VLBI prior (hereafter referred to as timing-only positions).

To test the PINPT strategy with pulsar timing, we re-derived $\left(\alpha^\mathcal{B}_2,\delta^\mathcal{B}_2\right)$ and $\left(\alpha^\mathcal{B}_3,\delta^\mathcal{B}_3\right)$ at $t_\mathrm{ref}=\mathrm{MJD}~55743$ (same as that of G21) following the method described earlier in Sect.~\ref{sec:results_and_implications}, under the assumption that the structures of J2221 and J2222 do not evolve with time. The results are displayed in Table~\ref{tab:abs_pos}. Intermediate results for the calculations are provided in Table~\ref{tab:pos_err_budget}.
Compared to the absolute pulsar position determined for MJD~59554 using either the J2221 or the J2222 approach, the pulsar position derived for MJD~55743 with the same approach is reported with smaller uncertainty. This is because the reference pulsar position $\Vec{x}_\mathrm{psr}^\mathrm{\,PR}$ is better constrained at MJD~55743 (see Appendix~\ref{subsec:pathway_thru_J2221} and Table~\ref{tab:pos_err_budget}), around which the pulsar was observed more densely with VLBI.

To minimize the correlation between the timing and VLBI positions, we only compare the VLBI positions to the two timing-only positions (of G21), which are listed in Table~\ref{tab:abs_pos} as the ``DMX'' and ``non-DMX'' positions. Here, ``DMX'', named by G21, refers to the pulsar timing model that describes DM with a piecewise constant function (\citealp{Demorest13}; G21), while "non-DMX" refers to the timing model that approximates DM variations with a cubic function (G21).
The DMX and non-DMX timing positions are marginally consistent to within $2\,\sigma$ (here and hereafter, $\sigma$ refers to the addition-of-quadrature of the uncertainties of the two compared sides), with the uncertainty of the DMX position being more conservative than the non-DMX one.

On the other side of the comparison, $\delta^\mathcal{B}_2$ and $\delta^\mathcal{B}_3$ are consistent with each other. However, $\alpha^\mathcal{B}_2$ is smaller than $\alpha^\mathcal{B}_3$ at $3\,\sigma$ significance, which is associated with a $\sim3\,\sigma$ discrepancy between the two $\mu_\alpha$ measured with respect to J2221 ($44.755^{+0.017}_{-0.014}$\,\maspy) and J2222 ($44.707\pm0.005$\,\maspy, as reported in Table~\ref{tab:astrometric_results}), respectively.
The discrepancy between $\alpha^\mathcal{B}_2$ and $\alpha^\mathcal{B}_3$ likely indicates the violation of the assumption that the structures of J2221 and J2222 do not evolve with time.
The jet direction of J2221 is almost aligned with the RA direction (see Fig.~\ref{fig:AGN_jet_directions}). As noted in D13, the relatively severe structure evolution of J2221 is believed to cause {\bf 1)} less precise parallax derived with respect to J2221 (as the parallax magnitude in the RA direction is $\approx2.4$ times larger than in the declination direction), and {\bf 2)} biased $\mu_\alpha$ determination. 
Specifically, the discrepancy between $\alpha^\mathcal{B}_2$ and $\alpha^\mathcal{B}_3$ (or the $\mu_\alpha$ discrepancy) can be explained by 0.56\,mas larger (or fractionally 18\% higher according to Table~\ref{tab:CS_results}) $r_\mathrm{J2221}\left(1.6\,\mathrm{GHz}\right)$ at MJD~55743 compared to at MJD~59554.
On the other hand, thanks to the horizontal jet of J2221, $\delta^\mathcal{B}_2$ is almost free from the impact of structure evolution (of J2221), hence being more favorable than $\delta^\mathcal{B}_3$, let alone the indication of vertical structure evolution of J2222 mentioned in Sect.~\ref{subsec:astrometric_inference}. Therefore, we adopt $\left(\alpha^\mathcal{B}_3,\delta^\mathcal{B}_2\right)$ as the absolute pulsar position at MJD~55743. 
A more sophisticated treatment in the future could incorporate positions in both axes from both sources, weighting them by some measure of expected reliability and taking into account the covariance between the two measurements.

When comparing $\left(\alpha^\mathcal{B}_3,\delta^\mathcal{B}_2\right)$ to the two timing-only positions, we first find good consistency between $\left(\alpha^\mathcal{B}_3,\delta^\mathcal{B}_2\right)$ and the DMX position, with $\alpha^\mathcal{B}_3$ and the DMX RA consistent to within $0.3\,\sigma$, and $\delta^\mathcal{B}_2$ and the DMX declination consistent to within $1.1\,\sigma$. On the other hand, despite the $1.5\,\sigma$ consistency between $\delta^\mathcal{B}_2$ and the non-DMX declination, $\alpha^\mathcal{B}_3$ is smaller than the non-DMX RA at $>3\,\sigma$ significance. 
It is not impossible that the $>3\,\sigma$ RA discrepancy is caused by the use of different kinds of reference systems.
Therefore, we cannot yet conclude that the new PINPT results favour the DMX timing model over the non-DMX one with just one MSP.

Moreover, we reiterate that $\alpha^\mathcal{B}_3$ is subject to additional errors induced by potential horizontal structure evolution of J2222. A more robust test of the PINPT strategy with pulsar timing can be achieved by comparing $\left(\alpha^\mathcal{B}_1,\delta^\mathcal{B}_1\right)$ to timing positions (based on different timing models) derived at MJD~59554. 
Preliminary timing analysis incorporating a large amount of new datasets shows reasonably good consistency with $\left(\alpha^\mathcal{B}_1,\delta^\mathcal{B}_1\right)$. Detailed results will be reported as part of an upcoming timing paper.
Additionally, applying the PINPT strategy to just a few well timed MSPs will allow a much stronger test of the new strategy.

\section{Summary}
\label{sec:implications}

Using the PINPT strategy, we determine an MSP position to the $\sim0.2$\,mas precision with VLBI, which improves on the previous precision level by a factor of $\sim5$ \citep[e.g.][]{Ding20,Liu23a}, and is comparable with the precision level of the timing positions of the best-timed MSPs \citep[e.g.][]{Perera19}. According to \citet{Liu23a}, applying the PINPT strategy to 50 MSPs promises $<0.1$\,mas precision for the connection between the kinematic and dynamic reference frames, $>3$ times more precise than the previous expectation. 
Considering that systematic errors of calibrator source positions are in a range of 0.05--0.2~mas\textsuperscript{\ref{footnote:rfc}} and unlikely to be improved within several decades,
our strategy provides the position accuracy that approaches to a practical limit.

In general, the PINPT strategy can be used in a broader context: it can sharpen the VLBI localisation of any steep-spectrum compact radio source, which could facilitate the studies of, for example, fast radio bursts \citep[e.g.][]{Bhandari23} or mergers of two neutron stars \citep[e.g.][]{Mooley22}.
Furthermore, it is important to stress that the PINPT strategy is not only meant to enhance the precision of absolute positions, but would have fundamental impact on differential astrometry as well. 
Due to the temporal structure evolution of the reference source(s) (which includes core-shift variations, \citealp{Plavin19}), proper motions and parallaxes measured with respect to AGNs could be potentially biased (e.g. D13, also see Sect.~\ref{subsec:comparison_VLBI_timing}), which is believed to have driven the occasional inconsistencies between VLBI and timing proper motions for the longest timed MSPs \citep{Ding23}. However, traditional differential astrometry with one calibrator is incapable of quantifying the positional variations due to the structure evolution, hence being subject to extra systematic errors. 
In the scenario of precise in-beam astrometry, variability in the core shifts of the reference sources is a leading source of systematic errors.
Multi-epoch PINPT observations, on the other hand, can constrain the core shift variabilities, thus minimizing their corruption on astrometric results. 
As pulsar astrometry is considered, joining the core-shift-determining sessions to in-beam or Multi-View astrometry can provide accurate pulsar proper motions and parallaxes not biased by core shift variabilities, which can then be safely incorporated into pulsar timing analysis.

\begin{acknowledgements}
HD appreciates the EACOA Fellowship awarded by the East Asia Core Observatories Association. HD thanks Wei Zhao for the discussion about jet direction determination. This work is mainly based on observations with the Very Long Baseline Array (VLBA), which is operated by the National Radio Astronomy Observatory (NRAO). The NRAO is a facility of the National Science Foundation operated under cooperative agreement by Associated Universities, Inc. Pulse ephemerides of \psr\ were made for the purpose of pulsar gating, using data from the Effelsberg 100-meter telescope of the Max-Planck-Institut f$\ddot u$r Radioastronomie. This work made use of the Swinburne University of Technology software correlator, developed as part of the Australian Major National Research Facilities Programme and operated under license \citep{Deller11a}. 

\end{acknowledgements}

%
%

\bibliographystyle{aa}
\bibliography{refs}

\begin{appendix} 

\section{Supporting materials for Sect.~\ref{subsec:obs}}
\label{appendixA}

\begin{figure}[h]
    \centering    \includegraphics[width=0.7\columnwidth]{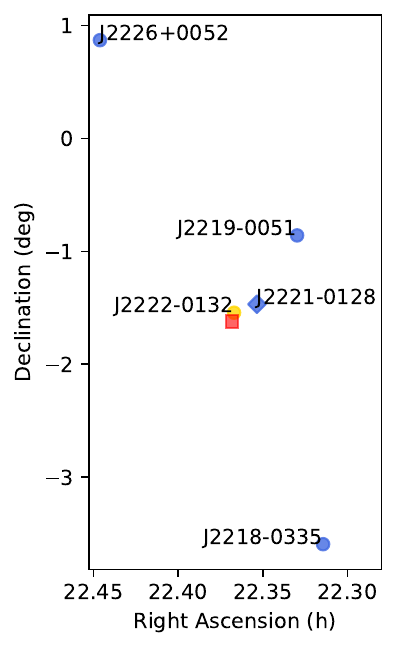}
    \caption{The calibrator plans for the PINPT observations of this work, which involve two in-beam phase calibrators and three out-of-beam calibrators of \psr\ (shown in red). 
    More source information can be found in Table~\ref{tab:sources}. Among the two in-beam calibrators, \ibcb, marked by golden circle, was used in previous astrometric campaigns \citep{Deller13,Guo21}; \ibca, marked by blue diamond, is used for core-shift determinations (see Sect.~\ref{subsubsec:CS_direction}). The adopted phase calibration strategies are described in Sect.~\ref{subsec:obs} and summarized in Table~\ref{tab:obs_summary}.}
    
    \label{fig:calibrator_plan}
\end{figure}

\begin{table*}[h]
     \centering
     \caption[]{\label{tab:sources}
     Unresolved flux densities  \& angular separations from \psr.
     
     }
     {\renewcommand{\arraystretch}{1.1}
     \resizebox{0.9\textwidth}{!}{
    \begin{tabular}{ccccccc}
        \hline 
        \hline
        source  & purpose \tablefootmark{a} & $\mathcal{S}_\mathrm{18cm}$ \tablefootmark{b} &  $\mathcal{S}_\mathrm{13cm}$ & $\mathcal{S}_\mathrm{4cm}$ & $\mathcal{S}_\mathrm{2cm}$ & $\Delta_\mathrm{psr}$ \\
       name   & & (\mjypb) & (\mjypb) & (\mjypb) & (\mjypb) & (deg) \\ 
        \hline
        
        \psr\   & target & 0.8  & ---  & --- & --- & 0 \\
        \agna\   & phase cal. / CS det. & $1.5\times10^3$  &  $1.0\times10^3$ & $1.4\times10^3$ & $1.8\times10^3$ & 2.13 \\
        \agnb\   & phase cal. / CS det. & 165.2  & 113.9  & 113.0  & 111.5 &  0.96 \\
        \agnc\   & phase cal. / CS det. & 249.2 &  267.7 & 459.3 & 762.2 &  2.75 \\
        \ibca\    & phase cal. / CS det. & 24.7 & 9.5  & 10.5 & 9.7 & 0.27 \\
        \ibcb\    & phase cal. & 14.5  &  --- & --- & --- & 0.08 \\
        \blz\  & fringe finder & $1.5\times10^3$ &  $1.1\times10^3$  & $1.5\times10^3$  & $2.2\times10^3$ & 12.07 \\
        \hline
        
        \hline
    \end{tabular}
    }}
    \tablefoot{$\mathcal{S}_\mathrm{X}$ refers to unresolved flux density at wavelength $X$ obtained from the BD244 data alone. $\Delta_\mathrm{psr}$ stands for angular separation from \psr.\\
    \tablefoottext{a}{``Phase cal.'' and ``CS det.'' refers to phase calibrator and core shift determination, respectively.\\}
    \tablefoottext{b}{The unresolved flux densities acquired from the first L-band observation, where the frequency band includes both $\sim1.44$\,GHz and $\sim1.76$\,GHz. The pulsar flux density is the average value over the spin period of \psr.}\\
    }
    \end{table*}

\section{Two approaches of Multi-View (2D interpolation)}
\label{appendix:2D_interpolation}

As mentioned in Sect.~\ref{subsubsec:2D_interpolation}, Multi-View was realized in two approaches. In both approaches, J2219 --- the closest to \psr\ among the off-beam  calibrators, was used as the main phase calibrator. 
In the first approach, we made a phase calibration with J2219, then interpolated the phase solution to the approximate sky position of \psr\ by performing two 1D interpolation operations \citep[e.g.][]{Ding20c} one after another. 
Specifically, the phase solution was first extrapolated along the J2219-to-J2226 line to the intersection with the straight line connecting J2218 and \psr. The derived phase solution was then extrapolated again along the J2218-to-pulsar line, to the position of \psr.

In the second approach, the phase solution acquired with J2219 was passed to J2218 and J2226. Subsequently, self-calibration was performed with both J2218 and J2226. The acquired incremental phase solutions $\phi_\mathrm{J2218}$ and $\phi_\mathrm{J2226}$ were then linearly added together as $c_\mathrm{J2218} \cdot \phi_\mathrm{J2218}+ c_\mathrm{J2226} \cdot \phi_\mathrm{J2226}$ (here, $c_\mathrm{J2218}$ and $c_\mathrm{J2226}$ are constants described in Appendix~\ref{appendix:deriving_abspos}), which essentially moves the virtual calibrator (explained in \citealp{Ding20c}) to the approximate sky position of the target (i.e., \psr, J2221, or J2222).

\section{$\hat{l}(\epsilon)$ and $\sigma_0(\nu)$}
\label{appendix:l0_and_sigma0}

In Eq.~\ref{eq:J2221_sysError}, there are two functions --- $\hat{l}(\epsilon)$ and $\sigma_0(\nu)$. The former describes the fractional atmospheric path length as a function of the antenna elevation $\epsilon$, while the latter characterizes the evolution of the systematic error with respect to the observing frequency $\nu$.
We define $\hat{l}(\epsilon)$ as $l/R_\mathrm{E}$, where $l$ and $R_\mathrm{E}$ are, respectively, the atmospheric path length and the radius of the Earth (not including the atmosphere). It is easy to calculate that 
\begin{equation}
\label{eq:fractional_path_length}
\hat{l}(\epsilon)=\overline{\sqrt{\sin^2{\epsilon}+2 \eta_\mathrm{H}+{\eta_\mathrm{H}}^2}-\sin{\epsilon}} \,\,,
\end{equation}
where the overline commands averaging over the observation; $\eta_\mathrm{H}=h_\mathrm{atmo}/R_\mathrm{E}$ is the atmosphere thickness divided by $R_\mathrm{E}$. We adopted $\eta_\mathrm{H}=0.15$ calculated with the upper height of the ionosphere ($\approx965$\,km) and the average Earth radius of 6371\,km.

Using simulations, the relation between propagation-related positional error and $\nu$ has been studied by \citet{Marti-Vidal10a}, and is provided in Fig.~6 of \citet{Marti-Vidal10a}.
As the simulations of \citet{Marti-Vidal10a} assume a 5\degr\ angular separation between the target and the reference source, we divided the results of \citet{Marti-Vidal10a} by a factor of 5 (to reach the systematics per degree separation), and adopted the divided results as $\sigma_0(\nu)$.
In Eq.~\ref{eq:J2221_sysError}, we assume $\sigma^\mathcal{S}_{ijk}\propto\hat{l}(\epsilon)\cdot\sigma_0(\nu_j)$, while having little knowledge about the coefficient. 
Therefore, the nuisance parameter $\eta_\mathrm{EFAC}$ (to be determined in Bayesian inference) is essential for completing Eq.~\ref{eq:J2221_sysError}, and recovering the true magnitude of $\sigma^\mathcal{S}_{ijk}$.

\section{The determination of AGN jet directions}
\label{appendixB:jet_direction_determination}

\begin{figure*}[hbt!]
    \begin{subfigure}{0.49\textwidth}
        \includegraphics[width=\textwidth]{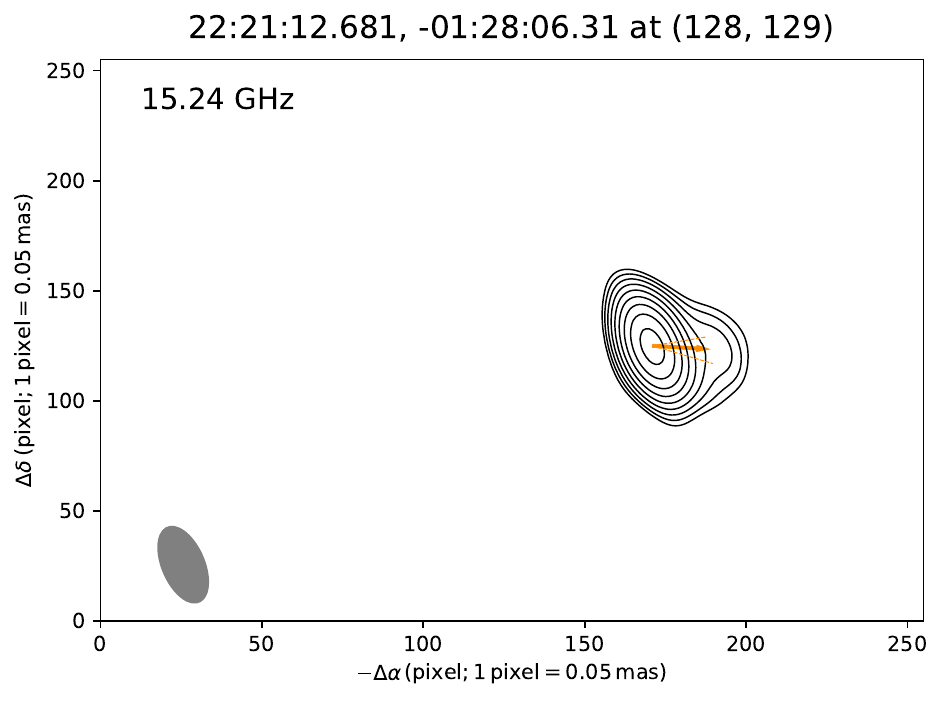}
        \caption[]%
            {\ibca}
    \end{subfigure}\hfill
    \vspace{5mm}
    \begin{subfigure}{0.49\textwidth}
        \includegraphics[width=\textwidth]{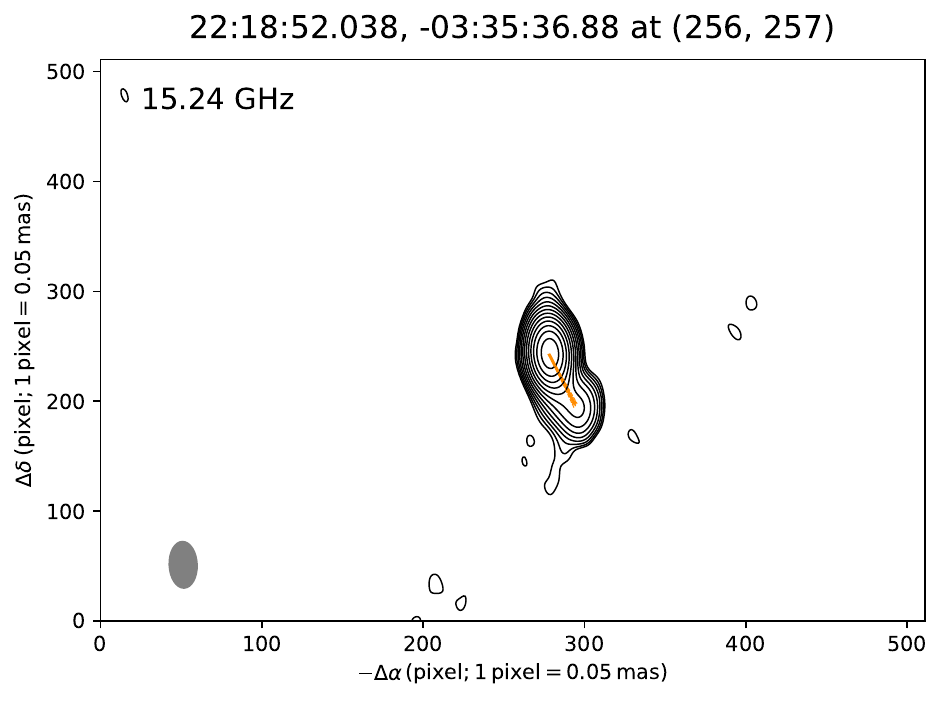}
        \caption[]%
            {\agna}
    \end{subfigure}

    \begin{subfigure}{0.49\textwidth}
        \includegraphics[width=\textwidth]{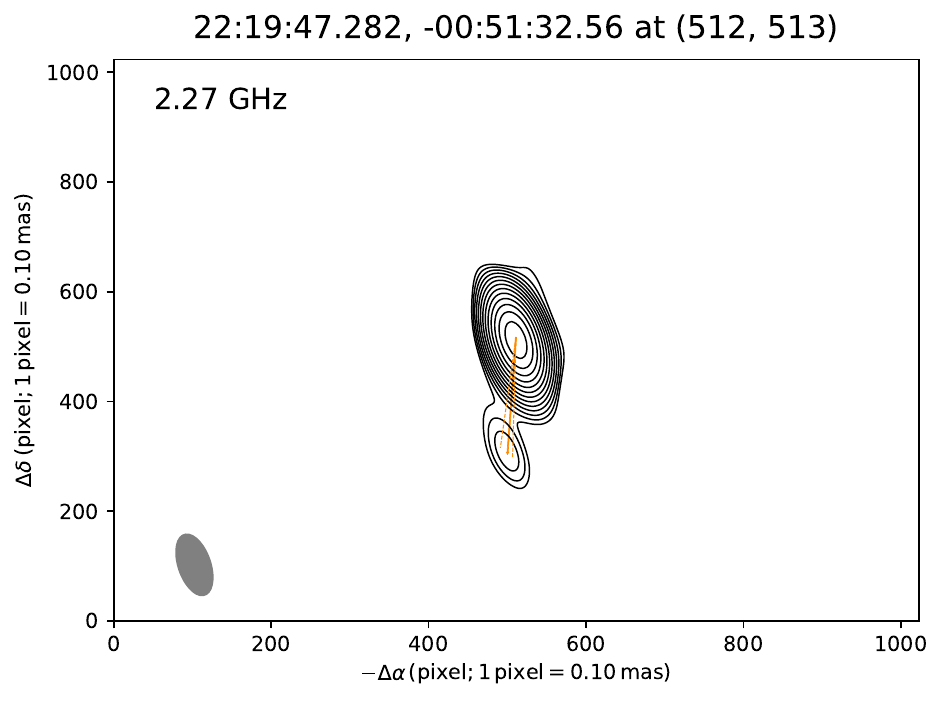}
         \caption[]%
            {\agnb}
    \end{subfigure}\hfill
    \begin{subfigure}{0.49\textwidth}
        \includegraphics[width=\textwidth]{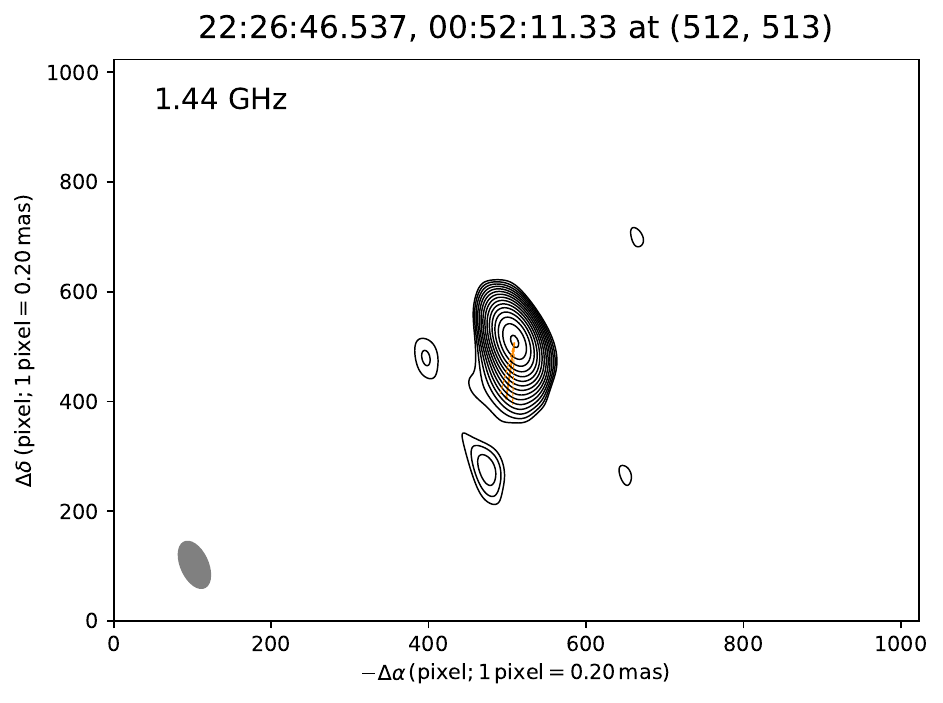}
        \caption[]%
            {\agnc}
    \end{subfigure}

    \caption[]
        {AGN contour plots obtained from the core-shift-determining sessions, and the AGN jet directions determined with {\tt arcfits}. 
        For the jet direction determinations, image models obtained at higher observing frequencies are preferred, unless extended jet features are not identified by {\tt arcfits}.
        The flux density on the $m$-th ($m=1, 2, 3, ...$) contour equals to $3 \cdot \mathrm{rms} \cdot \left(\sqrt{2}\right)^{m-1}$, where rms is 0.18\,\mjypb, 7.3\,\mjypb, 0.50\,\mjypb\ and 0.65\,\mjypb\ for J2221, J2218, J2219 and J2226, respectively.
        $\Delta\alpha$ and $\Delta\delta$ are, respectively, relative right ascension (left to the east) and relative declination in the unit of image pixel. The reference position of each image is provided over the image. In each panel, the gray ellipse shows the size of the synthesized beam; the AGN jet direction is marked with the orange arrow, while the uncertainty of the direction is illustrated with the two dashed lines.} 
        \label{fig:AGN_jet_directions}
\end{figure*}

AGN jet directions $\phi_\mathrm{jet}$ and their uncertainties have been directly estimated from their VLBI images \citep[e.g.][]{Kovalev17}. 
Likewise, we determined $\phi_\mathrm{jet}$ of the 4 core-shift-probing AGNs from the final image models\textsuperscript{\ref{footnote:image_models}} of the 4 sources using the newly developed package {\tt arcfits}\footnote{\label{footnote:arcfits}Available at \url{https://github.com/dingswin/arcfits}. The writing of the package has benefited from the discussion with Dr. Wei Zhao, who already has a preliminary script to derive directions along the jet ridge line using circular slicing (see \url{https://github.com/AXXE251/AGN-JET-RL}).}. 
For the $\phi_\mathrm{jet}$ determination, we prioritized the use of image models obtained at 15\,GHz because, compared to at lower frequencies, {\bf 1)} the jet core is closer to the central engine, and {\bf 2)} the spatial resolution is higher. In case extended jet features were not identified by {\tt arcfits}, we moved on to the image model of the next highest frequency.
The obtained $\phi_\mathrm{jet}$ results are illustrated alongside the AGN images in Fig.~\ref{fig:AGN_jet_directions}, and provided in Table~\ref{tab:jet_directions}.
{\tt arcfits} derives $\phi_\mathrm{jet}$ and their uncertainties in an automatic way described as follows.

\begin{table}[h]
     \centering
     \caption[]{\label{tab:jet_directions}
     AGN jet position angles $\phi_\mathrm{jet}$ determined with {\tt arcfits}.}
     {\renewcommand{\arraystretch}{1.4}
     \resizebox{0.6\columnwidth}{!}{
    \begin{tabular}{cc}
        \hline 
        \hline
        source name  & $\phi_\mathrm{jet}$ \\ 
          & (deg) \\ 
        \hline
        
        \agna\ & $220(3)$\\
        \agnb\ & $157^{+4}_{-5}$\\
        \ibca\ & $276^{+7}_{-9}$\\
        \agnc\ & $152^{+8}_{-7}$\\
        \hline
        
        \hline
    \end{tabular}
    }}
    \tablefoot{The adopted convention for the AGN jet position angles $\phi_\mathrm{jet}$ is east of north. The $\phi_\mathrm{jet}$ results are used as priors of the core shift directions in the Bayesian inference of core shifts.
    }
    \end{table}

\subsection{The noise level of the residual map}
\label{subsec:rms}

After reading a VLBI image, a 2D array of flux densities can be obtained.
From this 2D map, the noise level rms of the residual map (i.e., the image after removing all detected source components) can be derived, which is a preparation for measuring $\phi_\mathrm{jet}$. 
We established the residual map and estimated the rms in the following way. 
We first marginalized the 2D array of flux densities into a 1D array, and calculated the standard deviation of the flux densities. All flux densities higher than 7 times the standard deviation were considered ``detected'', and were removed from the 1D array. We repeated this standard deviation calculation and removal of detected points until we reached the residual map, in which no more detection can be identified. The standard deviation of this residual map was adopted as the residual map noise level rms.

\subsection{Elliptical cuts on the inner regions of VLBI images}
\label{subsec:elliptical_cuts}

The hitherto most advanced method of measuring the position angles of radio features outside the compact radio core is by 1) converting the positions in a VLBI image from Cartesian coordinate to a polar coordinate centered around the brightest spot of the image, and 2) applying circular cuts to the VLBI image \citep[e.g.][]{Cui23}.
As the core shift study is concerned, a $\phi_\mathrm{jet}$ determined close to the compact radio core is expected to better approximate the core shift direction, as compared to a $\phi_\mathrm{jet}$ determined afar \citep[e.g.][]{Konigl81}.
Due to intrinsic structure and/or scatter broadening, the compact radio core is usually larger than the synthesized beam, while normally resembling the beam (e.g., Fig.~\ref{fig:AGN_jet_directions}). 
In other words, there is no prior information about the size of the compact radio core.
As a result, when the synthesized beam is non-circular, circular cuts applied in close proximity to the compact radio core might lead to the misidentification of the compact radio core (stretching along the major axis) as radio features outside the compact core. Although this anomaly can be corrected by human intervention, other systematic errors might be introduced during the human intervention.

As a novel method that is {\bf i)} largely free of human intervention, and {\bf ii)} dedicated to $\phi_\mathrm{jet}$ determination in close proximity to the compact radio core, we cut an image model with a number of increasingly large ellipses that {\bf a)} have the same axis ratio and position angle as the synthetic beam, and {\bf b)} are centered at the image pixel of the highest flux density. Given that the position angle and axis ratio are both constant, the $n$-th ($n=1,2,3,...$) ellipse can be characterized by its semi-major axis $a_n$; a position on this ellipse can be defined with one additional parameter --- the position angle (east of north) $\phi$. In this work, we universally adopted 
\begin{equation}
\label{eq:a_n}
a_n=a_\mathrm{beam}+(n-1)\cdot \left(a_\mathrm{beam}/2\right)
\end{equation}
for the ellipses, where $a_\mathrm{beam}$ refers to the semi-major axis of the synthesized beam. 

Among the ellipses ascertained by Eq.~\ref{eq:a_n}, the innermost ellipse outside the compact radio core was used to derive $\phi_\mathrm{jet}$; the position angle corresponding to the maximum flux density on the ellipse was adopted as the $\phi_\mathrm{jet}$. 
An ellipse is considered outside the compact radio core, when it meets the following criteria:
\begin{enumerate}[label=(\roman*)]
    \item the maximum flux density on the ellipse is $>7$\,rms (see Appendix~\ref{subsec:rms} for the meaning and calculation of rms), and
    \item the median flux density on the ellipse is $<3$\,rms, and
    \item $\phi_\mathrm{jet}$ uncertainty can be calculated with the method detailed in Appendix~\ref{subsec:err_jet_direction}. 
\end{enumerate}
If no ellipse meets the criteria, the AGN is considered a compact source without extended radio features.

\subsection{The uncertainty on the AGN jet direction}
\label{subsec:err_jet_direction}

In Appendix~\ref{subsec:elliptical_cuts}, we adopted the position angle of the maximum flux density $\mathcal{S}_\mathrm{max}$ on the innermost elliptical cut outside the compact radio core as the $\phi_\mathrm{jet}$. Random noises in the image might distort the flux density distribution, and deviate the position of the maximum flux density on the ellipse. The degree of flux density change due to random noises is limited: the chance of large flux density changes due to random noises is lower than smaller flux density changes. Given a flux density drop $\Delta\mathcal{S}$ induced by random noises, one can estimate the chance of the drop by counting flux densities $<\bar{\mathcal{S}}_\mathrm{R}-\Delta\mathcal{S}$ in the residual map (see Appendix~\ref{subsec:rms} for explanation of the residual map), where $\bar{\mathcal{S}}_\mathrm{R}$ stands for the mean flux density of the residual map. Reversely, provided a confidence level where the flux densities in the residual map is $\geq\bar{\mathcal{S}}_\mathrm{R}-\Delta\mathcal{S}$, we can derive the $\Delta\mathcal{S}$ corresponding to the confidence level.
In this way, we estimated $\Delta\mathcal{S}$ corresponding to 68\% confidence level from the residual map.
At 68\% confidence, we expect $S'_\mathrm{max}>S_\mathrm{max}-\Delta\mathcal{S}$, where $S'_\mathrm{max}$ is the maximum flux density changed by random noises. 

On the ellipse (where $\phi_\mathrm{jet}$ is determined), we identify all position angles $\phi$ where the flux densities equal to $S_\mathrm{max}-\Delta\mathcal{S}$. When exactly two $\phi$ are acquired, the two $\phi$ are adopted as the $1\,\sigma$ uncertainty interval of $\phi_\mathrm{jet}$. Otherwise (which is rare), we consider the ellipse likely still intersects the compact radio core, and move onto the next ellipse further afield.

\section{The Bayesian inference of core shifts}
\label{appendix:Bayesian_inference_of_CSs}

We derived the core shift model with {\tt quartet}\footnote{planned to be released alongside a future catalog paper}, a newly developed package dedicated to inferring core-shift-related parameters in a Bayesian manner, which is explained as follows.

\subsection{Mathematical formalism}

Apart from the aforementioned parameters $r_i$, $\theta_i$, $\beta_i$ and $\eta_\mathrm{EFAC}$, the reference positions $x^*_{ik}$ are also required in the model of core shifts. For the Bayesian inference, the likelihood function is 
\begin{equation}
\label{eq:likehood_function}
P_\mathrm{CS} \propto \left(\prod_{i}\prod_j \prod_k\sigma_{ijk}\right)^{-1} \exp{\left[-\frac{1}{2} \sum_i \sum_j\sum_k\left(\frac{x_{ijk}-\tilde{x}_{ijk}}{\sigma_{ijk}}\right)^2\right]} \,,
\end{equation}
where $x_{ijk}$ and $\tilde{x}_{ijk}$ refer to, respectively, the observed and the modeled J2221 positions with respect to the $i$-th reference source at observing frequency $\nu_j$; the total positional uncertainty $\sigma_{ijk}$ is the addition-in-quadrature of random and systematic errors.
In Eq.~\ref{eq:likehood_function}, the modeled J2221 positions $\tilde{x}_{ijk}$ follow the relation
\begin{equation}
\label{eq:x_ijk}
\begin{split}
\tilde{x}_{ij\alpha} &= \left[r_\mathrm{J2221}(\nu_j) \cos{\theta_\mathrm{J2221}} - r_i(\nu_j) \cos{\theta_i} \right] + x^*_{i\alpha} \\
\tilde{x}_{ij\delta} &= \left[r_\mathrm{J2221}(\nu_j) \sin{\theta_\mathrm{J2221}} - r_i(\nu_j) \sin{\theta_i} \right] + x^*_{i\delta} \,,
\end{split} 
\end{equation}
where the formalism of $r_{i'}(\nu_j)$ is given by Eq.~\ref{eq:r_freq_relation}.

\subsection{Priors}

For the Bayesian inference, we adopted the following prior information:
\begin{enumerate}[label=(\roman*)]
    \item When the inference of $\beta_{i'}$ is requested, the prior constraints of $\beta_{i'}$ follow a uniform distribution between 0.3 and 3, which is denoted as $\beta_{i'}\sim \mathcal{U}\left(0.3,3\right)$. 
    \item $r_{0i'}\sim\mathcal{U}\left(0,5\right)$, where the unit is mas.
    \item $x^*_{ik}\sim\mathcal{U}\left(\overline{x_{ij}}|_k-3\sigma_{x_{ij}}|_k,\,\, \overline{x_{ij}}|_k+3\sigma_{x_{ij}}|_k\right)$, where $\overline{x_{ij}}|_k$ and $\sigma_{x_{ij}}|_k$ stand for the average and standard deviation of $x_{ijk}$ over all observing frequencies $j$; one lower limit and one upper limit are universally used for all observing frequencies. 
    \item $\eta_\mathrm{EFAC}\sim\mathcal{U}\left(0,20\right)$.
    \item The prior constraints on $\theta_{i'}$ follow Gaussian distributions characterized by the AGN jet directions $\phi_\mathrm{jet}$ in Table~\ref{tab:jet_directions}. Namely, $\theta_{i'}\sim\mathcal{G}\left(\phi_\mathrm{jet}|_{i'}, \sigma_{\phi_\mathrm{jet}}|_{i'}\right)$. In the case of asymmetric $\phi_\mathrm{jet}$ uncertainty, the larger side of the uncertainty is used as the $\sigma_{\phi_\mathrm{jet}}|_{i'}$.
\end{enumerate}

\begin{table*}[h]
     \centering
     \caption[]{\label{tab:CS_results}
     Core shift models derived at the reference frequency of 2.27\,GHz with Bayesian inference.}
     {\renewcommand{\arraystretch}{1.5}
     \resizebox{\textwidth}{!}{
    \begin{tabular}{c|c|c|c|c}
        \hline 
        \hline
        Model  & \multicolumn{4}{c}{AGN name} \\
        \cline{2-5}
        parameter  & \agna\ & \agnb\ & \agnc\ & \ibca\  \\ 
        \hline
         & \multicolumn{4}{c}{$\beta_{i'}$ inferred $\rightarrow \chi^2_\nu=1.9$}  \\
        \cline{2-5}
        $\beta_{i'}$ & $0.8^{+0.3}_{-0.2}$ & $1.0^{+0.9}_{-0.5}$ & $1.4^{+0.8}_{-0.6}$ & $0.73^{+0.22}_{-0.18}$ \\
        $r_{0i'}$ (mas) & $3.45^{+0.33}_{-0.36}$ & $0.70^{+0.34}_{-0.26}$ & $2.3^{+0.5}_{-0.4}$ & $2.19^{+0.32}_{-0.25}$ \\
        $\theta_{i'}$ (deg) & $218\pm3$ & $156\pm5$ & $149^{+5}_{-6}$ & $284\pm5$ \\
        $x^*_{i\alpha}$ & $22^\mathrm{h}21^\mathrm{m}12\fs680988\!\pm\!0.07\,\mathrm{mas}$  & $22^\mathrm{h}21^\mathrm{m}12\fs681004\!\pm\!0.04\,\mathrm{mas}$ & $22^\mathrm{h}21^\mathrm{m}12\fs681011\!\pm\!0.09\,\mathrm{mas}$ & --- \\
        $x^*_{i\delta}$ & $-01\degr28'06\farcs30895\!\pm\!0.15\,\mathrm{mas}$ & $-01\degr28'06\farcs30891\!\pm\!0.08\,\mathrm{mas}$ & $-01\degr28'06\farcs30852\!\pm\!0.17\,\mathrm{mas}$ & --- \\
        \cline{2-5}
        $\eta_\mathrm{EFAC}$ & \multicolumn{4}{c}{$5.0^{+1.4}_{-1.1}$} \\
        \hline
        \hline
        & \multicolumn{4}{c}{$\beta_{i'}\equiv1$ $\rightarrow \chi^2_\nu=1.2$}  \\
        \cline{2-5}
        $r_{0i'}$ (mas) & $3.25\pm0.27$ & $0.69^{+0.28}_{-0.29}$ & $2.3\pm0.4$ & $1.98^{+0.17}_{-0.16}$ \\
        $\theta_{i'}$ (deg) & $218\pm3$ & $157\pm5$ & $150\pm5$ & $284\pm6$ \\
        $x^*_{i\alpha}$ & $22^\mathrm{h}21^\mathrm{m}12\fs680989\!\pm\!0.06\,\mathrm{mas}$ & $22^\mathrm{h}21^\mathrm{m}12\fs681006\!\pm\!0.03\,\mathrm{mas}$ & $22^\mathrm{h}21^\mathrm{m}12\fs681011\!\pm\!0.07\,\mathrm{mas}$ & --- \\
        $x^*_{i\delta}$ & $-01\degr28'06\farcs30897\!\pm\!0.13\,\mathrm{mas}$ & $-01\degr28'06\farcs30891\!\pm\!0.07\,\mathrm{mas}$ & $-01\degr28'06\farcs30844\!\pm\!0.16\,\mathrm{mas}$ & --- \\
        \cline{2-5}
        $\eta_\mathrm{EFAC}$ & \multicolumn{4}{c}{$4.8^{+1.4}_{-1.0}$} \\
        \hline
        
        \hline
    \end{tabular}
    }}
    \tablefoot{The upper block shows 
 the results when the power-law indices $\beta_{i'}$ (defined in Eq.~\ref{eq:r_freq_relation}) are included in the Bayesian inference, while the lower block gives the results when $\beta_{i'}$ are fixed to 1. 
 Other model parameters include the systematics-correcting factor $\eta_\mathrm{EFAC}$, the core shift magnitudes $r_{0i'}$ (at the reference frequency $\nu_0$), and the reference positions $\left(x^*_{i\alpha},x^*_{i\delta}\right)$, defined in Eqs.~\ref{eq:J2221_sysError}, \ref{eq:r_freq_relation}, and \ref{eq:x_ijk}, respectively. $\chi^2_\nu$ denotes reduced chi-square of the inference, which is closer to unity when $\beta_{i'}\equiv1$. 
    }
    \end{table*}

\section{The calculation of an absolute pulsar position and its uncertainty}
\label{appendix:deriving_abspos}

The determination of the absolute position of \psr\ through the PINPT strategy relies on well determined absolute positions of the off-beam  calibrators. This work is based on the off-beam  calibrator positions $\vec{x}^\mathrm{\,RFC}_i$ reported in the 2024A release of the Radio Fundamental Catalogue\textsuperscript{\ref{footnote:rfc}} (RFC).
Generally speaking, for Multi-View carried out with 3 off-beam  calibrators, the relation between the target position $\Vec{x}_\mathrm{target}$ and the off-beam  calibrator positions $\Vec{x}_q$ ($q=1,2,3$) is
\begin{equation}
\label{eq:2D_interpolation_position_relation}
\Vec{x}_\mathrm{target} = \sum_q^3 c_q  \Vec{x}_q \,,
\end{equation}
where $\Vec{x}_\mathrm{target}$ is known to a precision good enough to guide the Multi-View; $c_q$ are coefficients that can be solved with the additional condition 
\begin{equation}
\label{eq:2D_interpolation_sum_eq_1}
\sum_q^3 c_q = 1 \,.
\end{equation}
Therefore, in Multi-View, offsets in the off-beam calibrator positions would lead to offset in the target position (refined with Multi-View), as 
\begin{equation}
\label{eq:2D_interpolation_Delta_position_relation}
\Delta\Vec{x}_\mathrm{target} = \sum_q^3 c_q  \cdot \Delta\Vec{x}_q \,,
\end{equation}
where $\Delta\Vec{x}_q$ result from {\bf 1)} core shifts of the off-beam calibrators, and {\bf 2)} inaccurate positions of the off-beam calibrator image models, and are calculated by
\begin{equation}
\label{eq:dx_q}
\Delta\Vec{x}_q = \left(\vec{x}^\mathrm{\,model}_q - \vec{x}^\mathrm{\,RFC}_q \right) + \left[\Vec{r}_q\left(\nu_\mathrm{psr}\right) - \vec{r}_q^\mathrm{\,RFC} \right]\,.
\end{equation}
Here, $\vec{x}^\mathrm{\,model}_q$ refers to the image model position of the $q$-th off-beam calibrator; $\Vec{r}_q\left(\nu_\mathrm{psr}\right)$ denotes the core shift of the $q$-th off-beam calibrator at the observing frequency of interest (which is 1.6\,GHz, the observing frequency of the pulsar sessions, for this work); $\vec{r}_q^\mathrm{\,RFC}$ is the residual core shift of the RFC position.

Specific to this work, $q=i=\mathrm{J2212}, \mathrm{J2219}, \mathrm{J2226}$.
$c_i$ for various targets are provided in Table~\ref{tab:c_i}.
According to \citet{Porcas09}, $\vec{r}_i^\mathrm{\,RFC}=0$, when {\bf 1)} $\vec{x}^\mathrm{\,RFC}_i$ is derived with group-delay astrometry (that removes the ionosphere-induced group delay with dual-band or multi-band geodetic observations), and {\bf 2)} $\beta_i=1$. Both conditions are met for this work: $\vec{x}^\mathrm{\,RFC}_i$ are estimated with group-delay astrometry; and we already assume that $\beta_i\equiv1$ in the Bayesian analysis (see Sect.~\ref{subsubsec:CS_freq_relation}). Therefore, we adopt $\vec{r}_i^\mathrm{\,RFC}=0$ in this work.

\begin{table}[h]
     \centering
     \caption[]{\label{tab:c_i}
     $c_i$ (of Eq.~\ref{eq:2D_interpolation_position_relation}) for different targets.}
     {\renewcommand{\arraystretch}{1.4}
     \resizebox{0.8\columnwidth}{!}{
    \begin{tabular}{cccc}
        \hline 
        \hline
        Target name & $c_\mathrm{J2218}$ & $c_\mathrm{J2219}$ & $c_\mathrm{J2226}$ \\ 
        \hline
        \psr\ & 0.53 & 0.07 & 0.40 \\
        \ibca\ & 0.38 & 0.36 & 0.25 \\
        \ibcb\ & 0.49 & 0.12 & 0.38 \\
        \hline
        
        \hline
    \end{tabular}
    }}
    \end{table}

The uncertainty on $\Delta\Vec{x}_\mathrm{target}$ is calculated by the addition-in-quadrature of the uncertainties on the components $c_i \Delta\vec{x}_i$, where the uncertainty of $\Delta\vec{x}_i$ is further derived by the addition-in-quadrature of the uncertainties on $\vec{x}^\mathrm{\,RFC}_i$ and $\Vec{r}_i$.

\subsection{Deriving an absolute pulsar position via a nearby AGN}
\label{subsec:pathway_thru_J2221}

When the target itself is an AGN (e.g. J2221, J2222), Eq.~\ref{eq:2D_interpolation_Delta_position_relation} needs to be generalized to 
\begin{equation}
\label{eq:2D_interpolation_Delta_position_relation_generalized}
\Delta\Vec{x}_\mathrm{target} = \sum_q^3 c_q  \cdot \Delta\Vec{x}_q + \Delta\Vec{r}_\mathrm{target} \,,
\end{equation}
where $\Delta\Vec{r}_\mathrm{target}$ is the core shift difference between $\nu_\mathrm{psr}$ and the observing frequency $\nu_\mathrm{IBC}$ of the AGN target.
Accordingly, the uncertainty on $\Delta\Vec{x}_\mathrm{target}$ is further added in quadrature by the uncertainty of $\Delta\Vec{r}_\mathrm{target}$.

The availability of a nearby AGN (hereafter referred to as IBC) around a pulsar (or other targets of interest) provides an alternative pathway to the absolute position of the pulsar. 
After applying the position correction $\Delta\Vec{x}_\mathrm{target}$ calculated with Eq.~\ref{eq:2D_interpolation_Delta_position_relation_generalized}, the absolute position $\vec{x}_\mathrm{IBC}$ of the IBC at $\nu_\mathrm{psr}$ can be derived.  Provided {\bf i)} the image model position $\Vec{x}_\mathrm{IBC}^\mathrm{\,model}$ of the IBC, and {\bf ii)} the pulsar position $\Vec{x}_\mathrm{psr}^\mathrm{\,PR}$ determined with phase referencing with respect to the IBC, the absolute pulsar position can be calculated as
\begin{equation}
\label{eq:abspos_via_IBC}
\Vec{x}_\mathrm{psr} = \Vec{x}_\mathrm{psr}^\mathrm{\,PR} + \left(\vec{x}_\mathrm{IBC} - \Vec{x}_\mathrm{IBC}^\mathrm{\,model} 
 \right) \,.
\end{equation}
This pathway of deriving the absolute position may become the only option in cases where Multi-View of the target of interest cannot be arranged (e.g., when the target is an unpredictable radio transient).
The uncertainty of $\Vec{x}_\mathrm{psr}$ derived via the IBC is the addition-in-quadrature of the $\Delta\Vec{x}_\mathrm{target}$ uncertainty, the $\Vec{x}_\mathrm{psr}^\mathrm{\,PR}$ uncertainty (estimated with Bayesian inference in this work), and the statistical uncertainty of the IBC position obtained with Multi-View at $\nu_\mathrm{IBC}$.

\begin{table*}[h]
     \centering
     \caption[]{\label{tab:pos_err_budget}
     Intermediate results for calculating the absolute positions reported in Table~\ref{tab:abs_pos}, and their associated uncertainties}
     {\renewcommand{\arraystretch}{1.5}
     \resizebox{\textwidth}{!}{
    \begin{tabular}{c|c|c}
        \hline 
        \hline
          & $\alpha^\mathcal{B}$ & $\delta^\mathcal{B}$ \\
\cline{2-3}
            & \multicolumn{2}{c}{$t_\mathrm{ref}=\mathrm{MJD}~59554.0$} \\
         \hline
        psr. appr. \tablefootmark{A} & $(22^\mathrm{h}22^\mathrm{m}05\fs99997\!\pm\!0.11\,\mathrm{mas})\!+\![(3.67\!\pm\!0.01)\!-\!(0.51\!\pm\!0.11)\!-\!(0.19\!\pm\!0.07)]\,\mathrm{mas}$ 
        & $(-01\degr37'15\farcs7825\!\pm\!0.25\,\mathrm{mas})\!+\![(1.39\!\pm\!0.01)\!-\!(1.90\!\pm\!0.19)\!+\!(0.34\!\pm\!0.08)]\,\mathrm{mas}$ \\
        J2221 appr. \tablefootmark{B} & $(22^\mathrm{h}22^\mathrm{m}06\fs00025\!\pm\!0.15\,\mathrm{mas})\!+\![(-0.53\!\pm\!0.01)\!-\!(1.43\!\pm\!0.10)\!+\!(0.43\!\pm\!0.06)]\,\mathrm{mas}$ 
        & $(-01\degr37'15\farcs7804\!\pm\!0.38\,\mathrm{mas})\!+\![(-2.69\!\pm\!0.02)\!+\!(0.04\!\pm\!0.16)\!+\!(0.58\!\pm\!0.08)]\,\mathrm{mas}$ \\
        J2222 appr. \tablefootmark{B} & $(22^\mathrm{h}22^\mathrm{m}06\fs00021\!\pm\!0.06\,\mathrm{mas})\!+\![(-0.22\!\pm\!0.03)\!-\!(0.45\!\pm\!0.11)\!-\!(0.18\!\pm\!0.07)]\,\mathrm{mas}$ 
        & $(-01\degr37'15\farcs7808\!\pm\!0.18\,\mathrm{mas})\!+\![(-0.54\!\pm\!0.06)\!-\!(1.82\!\pm\!0.18)\!+\!(0.34\!\pm\!0.08)]\,\mathrm{mas}$ \\

          \hline
        \hline
        
         & \multicolumn{2}{c}{$t_\mathrm{ref}=\mathrm{MJD}~55743$} \\
        \hline
        J2221 appr. \tablefootmark{B} & $(22^\mathrm{h}22^\mathrm{m}05\fs96911\!\pm\!0.06\,\mathrm{mas})\!+\![(-0.53\!\pm\!0.01)\!-\!(1.43\!\pm\!0.10)\!+\!(0.43\!\pm\!0.06)]\,\mathrm{mas}$ 
        &  $(-01\degr37'15\farcs7242\!\pm\!0.15\,\mathrm{mas})\!+\![(-2.69\!\pm\!0.02)\!+\!(0.04\!\pm\!0.16)\!+\!(0.58\!\pm\!0.08)]\,\mathrm{mas}$ \\
        J2222 appr. \tablefootmark{B} &  $(22^\mathrm{h}22^\mathrm{m}05\fs96910\!\pm\!0.01\,\mathrm{mas})\!+\![(-0.22\!\pm\!0.03)\!-\!(0.45\!\pm\!0.11)\!-\!(0.18\!\pm\!0.07)]\,\mathrm{mas}$
        & $(-01\degr37'15\farcs7244\!\pm\!0.03\,\mathrm{mas})\!+\![(-0.54\!\pm\!0.06)\!-\!(1.82\!\pm\!0.18)\!+\!(0.34\!\pm\!0.08)]\,\mathrm{mas}$ \\
    
        \hline
        
        \hline
    \end{tabular}
    }}
    \tablefoot{Identical to Table~\ref{tab:abs_pos}, the ``psr. appr.'', ``J2221 appr.'' and ``J2222 appr.'' refer to the three approaches (to the absolute position of \psr) described in Sect.~\ref{sec:results_and_implications}. Each absolute pulsar position is calculated from four terms in the round brackets. In all cases, the third term from the left refers to the core-shift-related position correction; and the fourth term is the position shift for aligning the off-beam calibrators to the RFC positions\textsuperscript{\ref{footnote:rfc}}.    
  \\
    \tablefoottext{A}{The first term from the left is the pulsar position obtained with the Multi-View strategy (see Sect.~\ref{subsubsec:2D_interpolation}), while the second term removes the parallax effects (see Sect.~\ref{sec:results_and_implications}).}\\
    \tablefoottext{B}{The first term from the left is the reference pulsar position $\Vec{x}_\mathrm{psr}^\mathrm{\,PR}$ derived with respect to the in-beam calibrator of use (i.e., J2221 for the J2221 approach, and J2222 for the J2222 approach)}. The second term aligns the reference point of the in-beam calibrator to the in-beam calibrator position determined with the Multi-View strategy (see Sect.~\ref{subsubsec:2D_interpolation}).}
    \end{table*}

\end{appendix}

\end{document}